\documentclass[
 superscriptaddress,
 aip, cha, 
 amsmath,amssymb,
 reprint,%
 author-year,
longbibliography
]{revtex4-2}
\usepackage[utf8]{inputenc}
\usepackage[english]{babel}
\usepackage{color} 
\usepackage{graphicx} 
\usepackage{wrapfig}
\usepackage{amsmath,amsthm,amsfonts,amssymb,amscd}
\usepackage{calc} 
\usepackage{array} 
\usepackage{titlesec} 
\usepackage{enumitem} 
\usepackage{changepage} 
\usepackage{hyperref} 
\usepackage{miller} 
\usepackage{soul} 
\usepackage{setspace} 
\usepackage{lipsum}
\usepackage{letltxmacro}  
\usepackage{float} 

\usepackage[defaultcolor=red]{changes}

\newcommand\mytitle{Optimizing Atomic Number Contrast in Multislice Electron Ptychography
}

\definecolor{linkColor}{rgb}{0.7,0,0}
\definecolor{darkred}{rgb}{0.7,0,0}

\hypersetup{
    pdftitle={\mytitle},
    pdfauthor={Colin Gilgenbach},
    pdfpagemode=FullScreen,
    pdfborder={0 0 0},
    colorlinks=false
}

\renewcommand{\AA}[0]{\text{\normalfont\r{A}}}

\titleformat{\section}
{\color{darkred}\sffamily\bfseries}
{\color{darkred}\thesection}{1em}{}
\titleformat{\subsection}
{\color{darkred}\sffamily\itshape}
{\color{darkred}\thesubsection}{1em}{}
\titlespacing*{\section}
{0pt}{8pt}{0pt}
\titlespacing*{\subsection}
{0pt}{8pt}{0pt}

\emergencystretch=\maxdimen
\LetLtxMacro{\ORIGselectlanguage}{\selectlanguage}
\makeatletter
\DeclareRobustCommand{\selectlanguage}[1]{%
  \@ifundefined{alias@\string#1}
    {\ORIGselectlanguage{#1}}
    {\begingroup\edef\x{\endgroup
       \noexpand\ORIGselectlanguage{\@nameuse{alias@#1}}}\x}%
}
\newcommand{\definelanguagealias}[2]{%
  \@namedef{alias@#1}{#2}%
}
\makeatother

\definelanguagealias{en}{english}
\definelanguagealias{EN}{english}
\definelanguagealias{de}{english}

\begin{document}

\title[\mytitle]{\mytitle}

\author{Bridget R. Denzer}
\affiliation{Massachusetts Institute of Technology}

\author{Colin Gilgenbach}
\affiliation{Massachusetts Institute of Technology}

\author{James M.~LeBeau}
\email{lebeau@mit.edu}
\affiliation{Massachusetts Institute of Technology}

\date{\today}

\begin{abstract}
    
    
    
    Here we explore the atomic number ($Z$) dependence of multislice electron ptychography and approaches to optimize Z sensitivity. Specifically, we show that ptychography's $Z$-dependence  is highly dependent on the integrated area of an atom column considered. A monotonic $Z$-dependence is found when the reconstructed projected atomic potentials are integrated over a small region. When increasing the integration area, $Z$-contrast changes significantly, becoming highly non-monotonic and following trends in the  orbital shell-structure. Moreover, the reconstructed projected potential aligns with the transmission function with an overall deviation of only 2.4\%. The non-monotonic $Z$-dependence is further shown to be useful to accentuate contrast between certain elements, allowing for distinguishability of elements that are only a single atomic number apart, and even in $>$ 20 nm thick samples. This is  demonstrated for $\beta$-CuZn ($Z$ = 29 and 30), with the differentiability between the elements explored for different signal quantification methods. The impact of electron dose and finite effective source size are also considered. These results demonstrate that the atom column integration area can optimize ptychographic $Z$-contrast for specific applications and experimental conditions.
    

    
\end{abstract}
\maketitle

\section{Introduction}

Imaging modalities that exhibit atomic-number ($Z$) contrast are indispensable tools in electron microscopy, as they allow one to infer local elemental occupancy. In high-angle annular dark-field (HAADF) scanning transmission electron microscopy (STEM), for example, $Z$-contrast enables analysis of composition at interfaces \citep{carlino_atomic-resolution_2005, browning_atomic-resolution_1993, falke_atomic_2004}, defects \citep{pennycook_chemically_1988, yan_impurity-induced_1998}, and chemical disorder \citep{yan_z-contrast_1998, yan_direct_1998}, which are critical to understanding the behavior of many materials. Beyond 2D materials, however, HAADF cannot easily resolve low-$Z$ elements like oxygen or nitrogen when heavier atoms are present \citep{yucelen_phase_2018, findlay_dynamics_2010, krivanek_atom-by-atom_2010}. Despite this limitation, HAADF is a widely-used $Z$-contrast technique due to its straightforward interpretation. 

Recently, multislice electron ptychography has gained interest as a robust phase retrieval technique that allows for imaging of both light and heavy elements with high spatial resolution. 
By collecting a 4D STEM dataset that consists of diffraction patterns sampled at a set of probe positions within a region of interest, the technique iteratively reconstructs the object projected potential and incident probe wavefunction \citep{yang_simultaneous_2016, chen_mixed-state_2020}. Further, incorporating incoherence and the multislice forward model has enabled deep sub-angstrom ($<\! 30$ pm) resolution \citep{chen_electron_2021, sha_deep_2022, jiang_electron_2018, chen_mixed-state_2020}. 
Through this approach, ptychography can largely remove the effects of dynamical scattering that are otherwise an inherent issue in other phase-contrast imaging techniques, e.g.~contrast reversals seen in high resolution TEM or DPC/COM STEM \citep{clark_effect_2023, yu_introduction_2024, lazic_phase_2016}. 



As a consequence of the enhanced resolution of multislice ptychography and roughly linear atomic number dependence of the projected potential, simultaneous imaging of light and heavy elements is straightforward even for relatively thick samples ($>\! 60$ nm) \citep{yang_imaging_2025, chen_electron_2021, chen_mixed-state_2020}. Moreover, the reconstructed projected potential from ptychography can be quantified, enabling direct analysis of the local element occupancy \citep{chen_electron_2021, hofer_reliable_2023}. 
 While \citet{chen_electron_2021} reported a $Z^{0.67}$ relation of multislice electron ptychography, understanding the $Z$-dependence as a function of analysis method, atomic number difference, dose, and incoherence is critical to enable routine analysis.



To quantify the projected potential, several methods to analyze conventional atomic-resolution STEM images can be considered, such as least-squares fitting the atom column intensities to a function, for example, 2D Gaussian or Lorentzian  \citep{hofer_atom-by-atom_2021, van_aert_quantitative_2009, hao_towards_2025}. With these approaches, either the atom column peak intensity or the atom column integrated intensity can be extracted. Further in HAADF imaging, integrating the signal over the atom columns has been shown to yield more robust and reliable measurements than using peak intensity values \citep{xia_haadfstem_2020, e_probe_2013, de_backer_atom_2013}. The improved resolution and information available with ptychography thus motivates the investigation of the interplay between the analysis approach and $Z$-dependence.

In this Article, we explore the influence of the analysis method on the $Z$-dependence of multislice electron ptychography reconstructed projected atomic potentials.  First, we consider a hypothetical structure with atom columns each comprised of a single element, from oxygen to francium ($Z$ = 8 to 87). Simulated 4D STEM datasets from this structure are then used to reconstruct the object projected potentials, which are measured with different approaches and compared. We show that the $Z$-dependence of these measurements strongly depends on the integration area and can be highly non-monotonic.  These results are further compared and contrasted with conventional imaging, namely HAADF and iDPC. Furthermore, we show that the $Z$-contrast between certain, neighboring elements can be significantly enhanced, e.g.~Cu and Zn, with appropriate limits of integration. The distinguishability of these elements is also explored as a function of electron dose and finite effective source size. Overall, these results show how atom column integration area can be selected to optimize $Z$-contrast in a particular material system. 




\section{Methods}

\subsection{Simulations}

A set of five `periodic table supercells' were created to explore $Z$-dependence of ptychography. Each $8.28 \times 8.28 \times 23.4$ nm supercell was constructed of tetragonal unit cells ($a = b = 3.345$ Å and $c = 5.45$ Å), where each block of four unit cells was a different element, as shown in Figure \ref{fig:potentials}a. To mark the central row and column of the supercell, polonium was used. A supercell of $\beta$-CuZn was also considered with a size of $7.36 \times 7.36 \times 20.0$ nm and lattice parameters of $a = b = c = 2.945$ Å \citep{gilat_normal_1965}.

Multislice 4D STEM simulations were performed following the method outlined by \citet{kirkland_advanced_2010}. The simulations considered a probe formed with 300 keV electrons, 15 nm overfocus, and a  convergence semi-angle ($\alpha$) of 25 mrad. Thermal diffuse scattering was incorporated using the frozen lattice approximation with 30 thermal configurations. All atoms in the periodic table supercells used the same Debye-Waller factor ($B$) of 0.50 Å$^2$ to separate  contributions to the $Z$-dependence from thermal vibration magnitude differences.   
Further, the selected Debye-Waller factor was representative of typical elemental crystal room-temperature thermal displacements  \citep{peng_debyewaller_1996, welberry_international_2021}. For CuZn, the average $B$-factor was calculated from the reported Debye temperature of 284 K \citep{shimizu_lattice_1976, warren_x-ray_1990}. Using $B_{Cu}$ approximately 12\% greater than $B_{Zn}$ based on \citet{chipman_long-range_1971}, the thermal displacements were based on $B_{Cu}$ = 0.69 Å$^2$ and $B_{Zn}$ = 0.62 Å$^2$.


The diffraction and scan step pixel sizes were 0.78 mrad/pixel and  0.56 Å/pixel, respectively. The finite size of the electron source was approximated by shifting the electron probe by a random amount at each real space pixel position in each thermal configuration, sampled from a Gaussian distribution with a full-width at half-maximum (FWHM) ranging from  0 to 0.75 Å.  Doses ranging from $5.9 \times 10^3 \:\mathrm{e^-/\AA^2}$ to $\infty$ were considered by adding corresponding levels of shot noise to the simulated 4D STEM patterns.


The HAADF inner and outer collection semi-angles were 70 and 200 mrad, respectively. A four-segment, annular detector was used to simulate DPC images with inner and outer collection semi-angles of 10.6 mrad and 28.2 mrad, respectively. While for HAADF the defocus was set to the entrance surface, the DPC simulations used a defocus of 8 nm (periodic table simulations) or 7 nm (CuZn), as determined by finding the maximum contrast value in a simulated defocus series. The finite effective source size \citep{NELLIST199461, lebeau_quantitative_2008, dwyer_method_2008, dwyer_measurement_2010} for the HAADF and iDPC simulations was included by convolving the images with a two-dimensional Gaussian with full-width at half-maxima of  0.25, 0.50, or 0.75 Å. 

\subsection{Reconstructions}

Multislice electron ptychography reconstructions were performed using the
\texttt{fold\_slice} fork of the PtychoShelves software package \citep{wakonig_ptychoshelves_2020, chen_electron_2021, odstrcil_iterative_2018, thibault_high-resolution_2008, thibault_maximum-likelihood_2012, thibault_reconstructing_2013, jiang_electron_2018}. All reconstructions used 8 probe modes and maximum collection semi-angle ($\beta$) of 50 mrad. The diffraction patterns were padded to 256 $\times$ 256 pixels in the last reconstruction engine.

\subsection{Quantification of Projected Potentials}

A residual background phase ramp of the reconstruction was removed by fitting a surface to the regions between the atom columns. The projected potential was determined from the reconstructed phase divided by the interaction parameter 
\citep{zou_phase_1999, fejes_approximations_1977, kirkland_advanced_2010}. The per-atom projected potential was determined by dividing the summed projected potential along the beam direction by the number of atoms in that atom column, such that the final units are projected in terms of V-Å/atom, i.e.~a sample thickness-independent quantity.

Two approaches to measure the projected potential at the atom columns. First, the atom columns were integrated over circular areas with radii of 0.20 Å, 0.40 Å, 0.64 Å, and 0.90 Å, which is referred to as the ``integration radius" below. This resulted in a per-atom integrated projected potential in units of V-Å$^3$/atom. Second, two-dimensional non-linear least squares fitting of Gaussians and Lorentzians was performed using the LMFIT python package \citep{newville_lmfit_2025}. 



\section{Results and Discussion}

\subsection{$Z$-dependence comparison}

\begin{figure}[htbp]
    \centering  
    \includegraphics[width=3.2in]{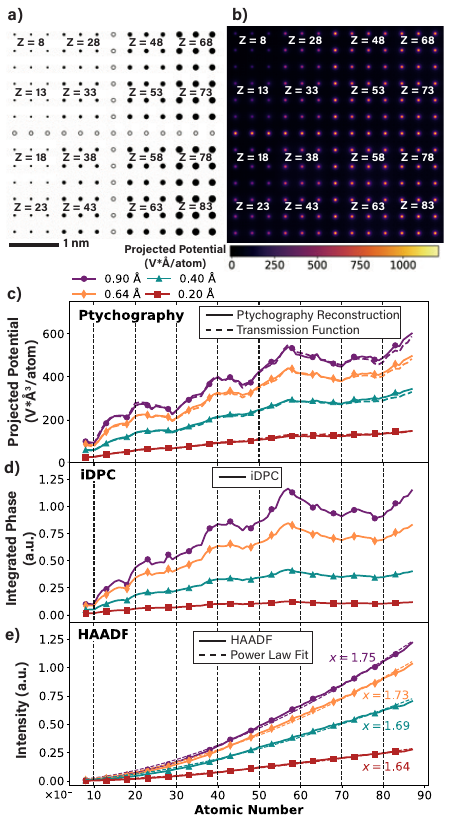}  
    \caption{(a) Schematic and (b) ptychography reconstruction of one of five supercells containing elements Z = 8 through Z = 87. Per-atom integrated projected potential using varying radii from the ptychography reconstruction are shown in (c) by the solid lines and the transmission function shown by the dashed lines. Similar analysis for (d) iDPC, and (e) HAADF along with the best fit $Z^x$ curves shown by dashed lines with $x$ listed on the plot. Markers are shown every 5 data points, and vertical dashed lines are shown every 10 data points.}
    \label{fig:potentials}  
\end{figure}

A representative schematic and reconstruction from one of the periodic table supercells is shown in Figure \ref{fig:potentials}a and \ref{fig:potentials}b, respectively. A clear contrast difference is observed between the light and heavy atoms in the reconstruction. The $Z$-dependence of the per-atom projected potentials is determined at each atom column for different radii, as shown Figure \ref{fig:potentials}c (solid lines) . Monotonic $Z$-dependence occurs for a small integration radius, e.g.~$r_0$ = 0.20 Å, with a  $Z^{0.66}$ relationship that is excellent agreement with \citet{chen_electron_2021}. 

When the integration radius is increased, e.g.~ $r_0$ = 0.90 Å, the $Z$-dependence is highly non-linear and instead follows trends in the electron orbital configuration and the filling of valence shells. For example, a local minimum is observed at the end of the first few periods (Z = 10, 18), as shown in Figure \ref{fig:potentials}c. This is followed by a decrease from Z = 21 to Z = 30 as the 3d subshell is filled and is consistent with the trends reported by \citet{kirkland_advanced_2010} for the projected potential root-mean-square radius. 

The observed shell-structure variation arises for larger integration radii because the projected potential contains not only the nuclear potential core but also a greater contribution from the outer valence shell electrons \citep{kirkland_advanced_2010}. As a result, integration using a smaller radius ($r_0$ = 0.20 Å) reflects a near-linear atomic-number scaling due to the nuclear potential, while larger-radius integration ($r_0$ = 0.90 Å) results in a shell structure Z-dependence.

The per-atom integrated projected potentials reconstructed from ptychography align closely with the transmission function, (dashed lines in Figure \ref{fig:potentials}c), with an average deviation of only 2.4\%. 
Furthermore, as sample thickness increases, the per-atom projected potential decreases, as shown in Supplementary Figure S1 for thicknesses of 10 nm, 16 nm, and 23 nm. The decrease is particularly noticeable for heavy atoms, possibly as a consequence of dynamical scattering to outside the detector range, and is 12.9\% on average.. Even with this decrease in signal, the overall trends in per-atom projected potential remain constant. Thus, the agreement of the reconstructed potential $Z$-dependence with the transmission function indicates that the trends are a reflection of the electrostatic potential.




Comparing ptychography to iDPC reveals the same shell-structure behavior at larger integration radii, as shown in  Figure \ref{fig:potentials}d. The observed non-monotonic $Z$-dependence is consistent with trends reported for center-of-mass (CoM) imaging shown by \citet{cao_theory_2018}. This is because iDPC, like CoM, is sensitive to the phase gradient, resulting in similar $Z$-dependence trends. However, there are a number of significant limitations with iDPC compared to ptychography. For example, while iDPC can be calculated in real-time, the contrast is highly sensitive to sample thickness, aberrations, and tilt \citep{calderon_accuracy_2022, burger_influence_2020}, making it difficult to achieve fully quantitative comparisons between experiment and theory. 

In contrast to iDPC, HAADF imaging does not produce the trends due to the electronic shell-structure, as shown in Figure \ref{fig:potentials}e. This is because the HAADF signal primarily captures Rutherford-like scattering from the nucleus, which follows a roughly $Z^{1.6-1.8}$ relation \citep{treacy_z_2011, pennycook_high-resolution_1991, hartel_conditions_1996, kirkland_advanced_2010}. 
The exponents in Figure \ref{fig:potentials}e are in agreement with these prior reports for thin samples ($\sim$5 nm). As thickness increases, the presence of heavier atoms in a 23 nm thick sample causes significant dynamical scattering \citep{yamashita_atomic_2018}, which decreases the exponent of the power-law fit. 
Thus, a well-characterized thickness for HAADF is critical for quantification \citep{lebeau_quantitative_2008}, whereas ptychography recovered projected potentials are nearly independent of thickness. 


\subsection{Optimizing Ptychography Z-contrast}\

 The shell-structure $Z$-dependence of multislice electron ptychography can enabled for optimized $Z$-contrast between elements that are close in atomic-number, similar to experimental results from monolayer hBN using single slice electron ptychography  \citep{martinez_direct_2019}. The occupancy of different valence electron subshells can create create large differences in the nuclear charge screening, leading to a larger variation in their per-atom integrated projected potentials at $r_0$ = 0.90 Å in Figure \ref{fig:potentials}, even for small $\Delta Z$. For example, consider chlorine ($Z= 17$) and potassium ($Z= 19$) or iodine ($Z = 53)$ and cesium ($Z = 55)$. This increased integrated projected potential difference would thus yield increased distinguishability. 

\begin{figure}[htbp] 
    \centering
    \includegraphics[width=3.2in]{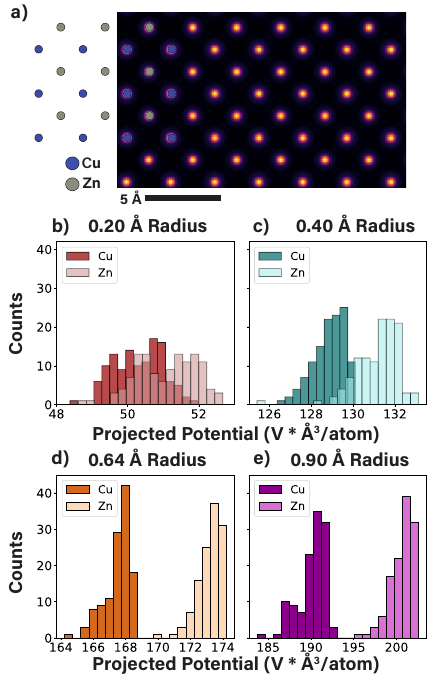}
    \caption{(a) Multislice electron ptychography reconstruction of CuZn. 
    Histograms of Cu and Zn atom columns integrated over $r_0$ = (b) 0.20 Å, (c) 0.40 Å, (d) 0.64 Å, and (e) 0.90 Å at a finite dose of 1.2 $\times$ 10$^6$ e/Å$^2$ and a finite source size of 0.25 Å.}
    \label{fig:cuzn_fig2}
\end{figure}

Making use of this shell structure can even be used to distinguish between elements with $\Delta Z$=1 in certain cases, e.g.~, Cu ($Z=29$) and Zn ($Z=30$). While difficult to visually distinguish, Figure \ref{fig:cuzn_fig2}a, the contrast can be measured using $\frac{2(I_{Cu}-I_{Zn})}{I_{Cu}+I_{Zn}}$, where $I_{Cu/Zn}$ are the measured integrated phase for Cu and Zn. As shown in Figure \ref{fig:cuzn_fig2}b-e, their contrast changes considerably with $r_0$ and  and has a critical impact on distinguishability. For example, when $r_0$ is small, 0.20 Å, the Cu and Zn distributions completely overlap and are thus indistinguishable. As $r_0$ increases, however, the distributions of the per-atom projected potentials become increasingly separated and are completely distinguishable when $r_0 > 0.64$ Å. At $r_0= 0.90$  Å, the contrast between Cu and Zn positions is 5.4\%, leading to complete distinguishability.




\subsection{Dose and Source Incoherence}

Dose and and incoherence also affect the distinguishability of similar-$Z$ elements in electron ptychography. While dose is governed by the probe current and dwell time, incoherence has several origins including thermal vibrations, the detector point spread function, and the finite effective source size to name a few. At both low dose and high incoherence, the reconstruction quality is significantly degraded \citep{chen_mixed-state_2020, li_atomically_2025}, making it increasingly difficult to identify similar-$Z$ elements.


To study the effects of dose and incoherence, simulations of CuZn were carried out using several finite source sizes, and the reconstructions were performed with finite dose. The contrast-to-noise ratio (CNR) is defined as:

\begin{equation}
    \frac{|\mu_{Cu} - \mu_{Zn}|}{\sqrt{\sigma_{Cu}^2+\sigma_{Zn}^2}}
\end{equation}

\noindent where $\mu$ and $\sigma$ are the respective average and standard deviation of the measured phase or intensity distributions, respectively.
\citep{timischl_contrastnoise_2015, celik_contrast_2002}. A larger CNR thus represents a larger separation of the per-atom integrated projected potentials compared to their standard deviations, and is thus used to determine Cu and Zn distinguishability 

\begin{figure*}[t]  
    \centering  
    \includegraphics[width=6.2in]{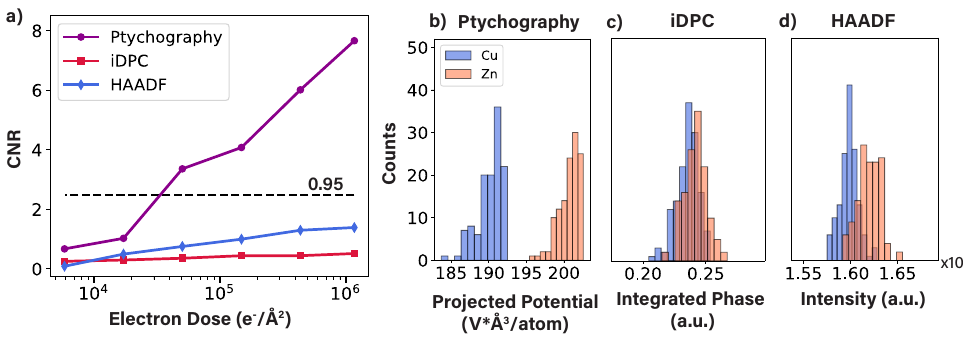}  
    \caption{(a) Contrast-to-noise ratios of simulated CuZn using ptychography, iDPC, and HAADF with a finite source size of 0.50 Å and varying electron dose. The black dashed line shows the threshold for 95\% distinguishability. The histograms of ptychography, iDPC, and HAADF are shown in (b), (c), and (d), respectively at a finite source size of 0.50 Å and dose of 1.2 $\times$ 10$^6$ e/Å$^2$.}
    \label{fig:cnr_ptycho_idpc_haadf}  
\end{figure*}

For a Gaussian finite source size of 0.50 Å FWHM, comparable to the measured effective source size for a Schottky emitter \citep{dwyer_method_2008},  the CNR for ptychography, iDPC, and HAADF are shown in Figure \ref{fig:cnr_ptycho_idpc_haadf}a. The threshold for 95\% true positive atom column identification, as shown by the black dashed line, is determined through the empirical cumulative distribution function.
While ptychography achieves $>$ 95\% accuracy for all doses except the lowest two, 5.9 $\times$ 10$^3$ and 1.7 $\times$ 10$^4$ e/Å$^2$, both iDPC and HAADF are below 83\% for the entirety of the dose range, indicating poor Cu/Zn atom column distinguishability. This is also clear from the 1.2 $\times$ 10$^6$ e/Å$^2$ dose case, Figure \ref{fig:cnr_ptycho_idpc_haadf}b-d, where the Cu and Zn atom column  distributions in both HAADF and iDPC exhibit significant overlap, while in ptychography the sublattices are readily separated. Thus, ptychography yields improved distinguishability over iDPC and HAADF for certain similar-Z elements, even in the presence of finite dose and finite source size.

\subsection{Integration method}

\begin{figure}[H]  
    \centering  
    \includegraphics[width=3.2in]{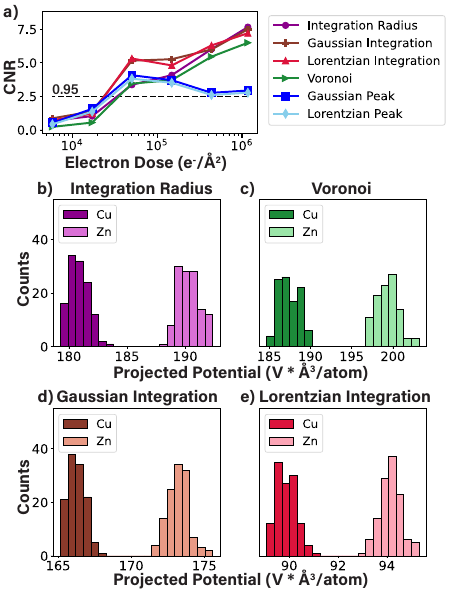}  
    \caption{(a) Contrast-to-noise ratio of ptychographic reconstructions quantified using different signal collection methods. The CNR values are shown for varying electron dose and a fixed finite source size of 0.50 Å. (b-e) Histograms of the collected per-atom projected potentials from Cu and Zn atoms at a dose of 1.2 $\times$ 10$^6$ e$^-$/Å$^2$ using a fixed integration radius of 0.90 Å, the Voronoi method, Gaussian integration, and Lorentzian integration.}
    \label{fig:voronoi}  
\end{figure}

To explore the ptychography distinguishability of Cu and Zn when using different signal quantification methods, the CNR is determined from measurements that integrate within a fixed radius as discussed above, Voronoi integration \citep{e_probe_2013}, and fitting the atom columns with a Gaussian and/or Lorentzian function(s), as shown in Figure \ref{fig:voronoi}a. To align with the fixed integration radius method, the Gaussian and Lorentzian functions are also integrated to a fixed radius of 0.90 Å. Both the integrated and peak values are considered for the Gaussian and Lorentzian fitting.

While all of the quantification methods considered result in high Cu/Zn distinguishability ($>95$\%) for doses $\geq 5.0 \times$ 10$^4$ e$^-$/Å$^2$, the integration methods provde higher CNR compared to the peak measurements for doses $\geq 1.5 \times$ 10$^5$ e$^-$/Å$^2$.
This indicates that increasing the integrated area is more important than the specific integration method; with sufficient dose, integrating is far better for distinguishability than using the peak alone \citep{e_probe_2013}.
For instance, at the highest dose of $1.2 \times 10^6$ e$^-$/Å$^2$, the CNR for the integrated methods is high (6.4 - 7.6), and the histograms show clear distinguishability between the Cu and Zn atom columns (Figure \ref{fig:voronoi}b-e).

\begin{figure*}[t]  
    \centering
    \includegraphics[width=6.2in]{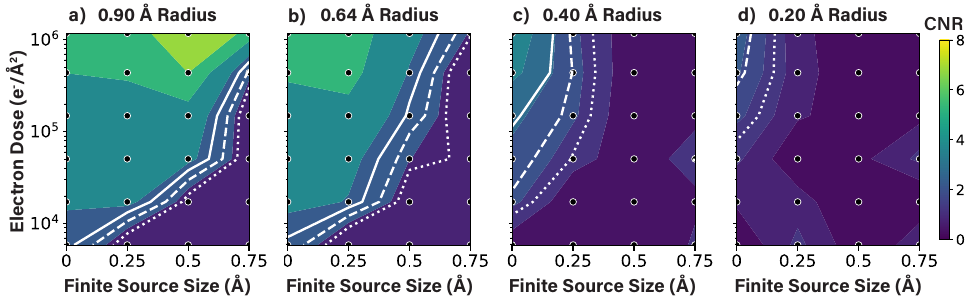}
    \caption{Contrast-to-noise ratio of CuZn reconstructions with varying electron dose and finite source size with different integration radii of (a) 0.90 Å, (b) 0.64 Å, (c) 0.40 Å, and (d) 0.20 Å. The white solid, dashed, and dotted lines show the thresholds for correctly identifying 95\%, 90\% and 80\% of the atom columns, respectively. Data points are shown by black markers.}
    \label{fig:cnr_cuzn}
\end{figure*}

The CNR dependence on integration area, dose, and finite source size is investigated further in Figure \ref{fig:cnr_cuzn}. Over a dose range of 5.9 $\times$ 10$^3$ -- 1.2 $\times$ 10$^6$ e$^-$/Å$^2$ and finite source size range of 0 -- 0.75 Å FWHM, at large $r_0$ (0.90 Å) much of the parameter space exhibits better than 95\% correct atom column identification (threshold indicated by solid white line). Reduced distinguishability occurs for low electron doses and high finite source sizes. This is because Poisson noise has a higher impact on the reconstruction quality at lower doses \citep{jiang_electron_2018}, resulting in increased variation, i.e.~increased $\sigma_\mathrm{Cu}$ and $\sigma_{\mathrm{Zn}}$, of the per-atom projected potential that reduces the CNR. 

When decreasing $r_0$ from 0.90 Å to 0.64 Å, the CNR decreases slightly as well, with the average CNR dropping from 3.5 to 2.8 (by 20\%). Even so, there is still a region in which the atom columns are 95\% distinguishable, although requiring slightly lower finite source size and higher dose, Figure \ref{fig:cnr_cuzn}b. Thus, when decreasing $r_0$, there is a stricter threshold for distinguishability of similar-Z elements, and a lower finite source size or higher dose is required to decrease the effects of experimental noise.

Below $r_0 = 0.40$ A, the CNR drastically decreases, as seen by Figure \ref{fig:cnr_cuzn}c and \ref{fig:cnr_cuzn}d. At $r_0=$ of 0.20 Å, for example, the atom columns are less then 80\% identifiable at most values in the parameter space. Only at a high dose ($> 4.4 \times 10^5$ e$^-$/Å$^2$) and a point source (source FWHM = 0) does the distinguishability reach 90\%. These parameters, however, are unreasonable to achieve in an experimental setting. The resulting poor distinguishability ($<$ 80\%) at reasonable experimental conditions does not provide sufficient $Z$-contrast to successfully differentiate between these atoms. 

Thus, analysis of projected potentials reconstructed by ptychography can be strategically optimized to enable improved differentiability of elements with similar atomic numbers, particularly in the presence of source incoherence and a particular dose.

\section{Conclusion}

The $Z$-dependence of multislice electron ptychography reconstructed projected potentials depends on the integrated atom column area. When using a small integration radius or peak phase, the $Z$-dependence follows, to a good approximation, a power law ($Z^{0.66}$). The dependence, however, becomes increasingly non-monotonic as the integration area increases, and follows trends in electron orbital shell structure, aligning with the transmission function to within 2.4\% on average. Further, the shell structure sensitivity can enable increased $Z$-contrast between certain similar-$Z$ elements, e.g. Cu and Zn even for a 20 nm thick sample. At a finite source size of 0.50 Å, ptychography achieves $>$ 95\% distinguishability between Cu and Zn for doses $>$ 1.7 $\times$ 10$^4$ e/Å$^2$, while both iDPC and HAADF remain below 83\% for all doses considered (up to 1.2 $\times$ 10$^6$ e/Å$^2$). Moreover, decreasing the ptychography integration radius significantly below 0.90 Å results in greatly diminished signal contrast and distinguishability except at a high dose ($> 4.4 \times 10^5$ e$^-$/Å$^2$) and with a point source. Thus, there is a stricter threshold for dose and finite source size when integrating with a smaller radius. 
Finally, these results demonstrate that with appropriate conditions and analysis, the $Z$-contrast capabilities of multislice electron ptychography can be optimized, which is crucial for extracting correlations between local structure and chemistry.




\bibliographystyle{MandM}

\bibliography{refs_static}


\end{document}